\documentclass{article}

    \PassOptionsToPackage{numbers, compress}{natbib}

\usepackage[preprint]{neurips_2025}

\usepackage[utf8]{inputenc} 
\usepackage[T1]{fontenc}    
\usepackage{hyperref}       
\usepackage{url}            
\usepackage{booktabs}       
\usepackage{amsfonts}       
\usepackage{amsmath}       
\usepackage{nicefrac}       
\usepackage{microtype}      
\usepackage{xcolor}         

\usepackage{graphicx} 
\usepackage{amsmath, amssymb, algorithm, algpseudocode}
\usepackage{algorithmicx}
\usepackage{algpseudocode}
\usepackage{algorithm}
\usepackage{booktabs}
\usepackage{multirow} 
\usepackage{enumitem}
\usepackage{subcaption}
\usepackage{titletoc}
\usepackage{caption}  

\usepackage{wrapfig}

\title{Robust and Efficient Autoregressive Speech Synthesis with Dynamic Chunk-wise Prediction Policy}

\author{%
  Bohan Li$^\ast$\quad
  Zhihan Li$^\ast$\quad
  Haoran Wang\quad 
  Hanglei Zhang\quad 
  Yiwei Guo\quad 
  Hankun Wang \\[0.04cm]
  \textbf{Xie Chen}\quad
  \textbf{Kai Yu}$^\dagger$ \\
    X-LANCE Lab, School of Computer Science; \\
    MoE Key Lab of Artificial Intelligence, AI Institute, \\
    Shanghai Jiao Tong University, Shanghai, China; \\
        MoE Key Lab of Artificial Intelligence,
        Jiangsu Key Lab of Language Computing \\
  \texttt{\{everlastingnight, kai.yu\}@sjtu.edu.cn} \\
}

\newcommand{\MethodName}{DCAR}

\begin{document}

\maketitle

\renewcommand{\thefootnote}{\fnsymbol{footnote}}
\setcounter{footnote}{0}
\footnotetext{$^\ast$Equal contribution.}
\footnotetext{$^\dagger$The corresponding author.}
\renewcommand{\thefootnote}{\arabic{footnote}}

\begin{abstract}
  Recently, autoregressive (AR) language models have emerged as a dominant approach in speech synthesis, offering expressive generation and scalable training. However, conventional AR speech synthesis models relying on the \textit{next-token prediction} paradigm often encounter significant challenges when handling long speech sequences. These models often struggle to construct stable frame-to-frame attention, leading to increased latency and degraded synthesis quality, thereby limiting their feasibility for real-time applications. To address these limitations, we introduce a novel \textbf{d}ynamic \textbf{c}hunk-wise \textbf{a}uto\textbf{r}egressive synthesis framework, termed \textbf{\MethodName}, designed to enhance both efficiency and intelligibility robustness in AR speech generation. \textit{\MethodName} introduces a chunk-to-frame attention mechanism through training with \textit{multi-token prediction}, enabling dynamic chunk prediction in variable speech contexts using a lightweight module trained on-policy. \textit{\MethodName} dynamically adjusts the token prediction span, significantly reducing the sequence length dependency while obtaining high synthesis quality. Comprehensive empirical evaluations demonstrate that \textit{\MethodName} substantially outperforms traditional \textit{next-token prediction} models, achieving up to \textbf{72.27\%} intelligibility improvement and \textbf{2.61$\times$} inference speedup \textbf{simultaneously} on test set. Furthermore, we conduct comprehensive analysis to support it as a versatile foundation for next-generation speech synthesis systems.
\end{abstract}

\vspace{-12pt}
\section{Introduction}
\vspace{-5pt}
Autoregressive (AR) architectures are widely employed in natural language processing (NLP) for both understanding and generation tasks. These models typically operate under the \textit{next-token prediction} paradigm, wherein each subsequent token is predicted based on the sequence of preceding tokens. This approach has underpinned many of the recent advances in text generation, particularly with the rise of large language models (LLMs)~\cite{zhao2023survey}.

In the speech domain, adopting similar AR modeling strategies naturally requires the use of quantization techniques to convert continuous acoustic signals into discrete representations, which effectively serve as a speech tokenizer. To this end, researchers have actively explored various discretization methods of speech~\cite{wu2024codec,guo2025recent,arora2025landscape}. Prominent approaches include applying vector quantization (VQ) on existing speech neural representation and training speech codecs using VQ-Variational Autoencoder (VQ-VAE)~\cite{VQVAE} architectures. Unlike textual or visual tokens, speech tokens typically correspond to short frames or time-windowed segments of audio, capturing low-level acoustic patterns but do not directly encode semantic information. These properties pose challenges for Transformer architectures, as lengthy speech sequences demand autoregressive modeling of long-range dependencies across frames using the \textit{next-token prediction} paradigm.

In this work, we investigate \textit{whether the frame-level autoregressive(\textit{FAR}) prediction paradigm is essential for autoregressive speech synthesis models}. Inspired by recent developments in LLMs, \textit{muti-token prediction}, which appears as a method for LLM fast decoding and rapid convergence in training, offers speech synthesis a promising direction to reduce the computational burden associated with lengthy speech sequences in AR models. 
We conduct an experimental study on autoregressive TTS using various configurations of speech tokens, revealing that tokens predicted by the first head \textbf{do not} consistently outperform those predicted by subsequent heads at the same sequence position. This phenomenon appears as a reasonable explanation for the robust synthesis by directly predicting without any speculative strategies a chunk of speech tokens at each decoding step. We refer to this method as \textit{chunk-wise autoregressive} synthesis, termed \textbf{\textit{CAR}}. While prior efforts~\cite{vadusa, nguyen2024acceleratingcodecbasedspeechsynthesis, vocalnet} have applied \textit{CAR} in autoregressive TTS using decoding techniques to achieve fast or high-quality synthesis, they still rely on parallel speculative verification based on the first head or simply accepting a fixed-size chunk of tokens. This leaves room for improvement in token selection strategies, as current approaches may retain redundant or suboptimal predictions. 

To this end, we propose \textbf{\textit{\MethodName}}, a dynamic chunk-wise synthesis strategy that uses two-stage training, allowing to schedule a proper chunk size at each \textit{CAR} decoding step. In the first stage, we let the TTS base model learn a chunk-to-frame attention pattern through \textit{muti-token prediction}, where the model predicts next $n$ tokens through one base head and $n-1$ additional heads. With the base head for the next token prediction, additional heads are responsible for predicting the second to $n$-th next succeeding tokens.
The losses for each head are summed with coefficients to form the final \textit{CAR} training criterion.

Sequentially, \textit{\MethodName} trains a lightweight policy based on the pre-trained TTS model to dynamically determine the length of next chunk through an on-policy strategy adapted from \textit{Group Relative Preference Optimization} \citep{shao2024deepseekmath}. Recent researches~\cite{chu2025sftmemorizesrlgeneralizes, yue2025doesreinforcementlearningreally} reveal that preference optimization policies prefer to let the model learn output preference rather than brand new knowledge, which means it is essential to give a guidance to the chunk-schedule policy, reducing the preference learning space. Consequently, we propose a \textit{chase-then-exceed} strategy, where we profile a range of chunk size that perform well in fixed-size \textit{CAR} and set them as half of the sampling policy in a group as guidance for the other sampling from the lightweight module trained on policy. Meanwhile, we give a negative process reward as a trace penalty for each outrange chunk size to further encourage the policy taking actions in our desired range. 
\textit{\MethodName} policy for chunk size schedule shows effectiveness through only two-epoch training on a tiny set with 980 samples from the LibriTTS~\cite{libritts} training set, conducting TTS with both stability and competitive speed.

The evaluation of \textit{\MethodName} is basically divided into four aspects: intelligibility robustness, naturalness of synthesis, speaker similarity of zero-shot 
 TTS and autoregressive decoding acceleration. \textit{\MethodName} simultaneously achieves up to 72.27\% (WER 9.99\% to 2.77\%) and 20.34\% (WER 2.31\% to 1.84\%) synthesis intelligibility improvement, 2.61$\times$ and 2.89$\times$ synthesis speedup TTS utilizing HuBERT~\cite{hsu2021hubert} and $\mathcal{S}^3$-Tokenizer~\cite{du2024cosyvoice} tokens.

Our main contributions can be summarized as follows:

\begin{itemize}[topsep=0pt, partopsep=0pt, leftmargin=7mm]
    \setlength{\itemsep}{2pt}
    \setlength{\parskip}{0pt}
    \item We propose \textit{\MethodName}, a novel autoregressive method for speech synthesis, which predict dynamic chunk-wise speech tokens autoregressively, benefiting synthesis speed and robustness. 
    
    \item We conduct lightweight policy training with \textit{DCPO} algorithm, which utilizes a \textit{chase-then-exceed} strategy. It effectively accelerates the convergence process and leverages well-performed interim results in a sufficient way.
    \item We analyze the \textit{chunk-to-frame} attention pattern and predicting behavior in the \textit{CAR} models, providing valuable insights in the nature of speech sequences self-attention.
    
\end{itemize}
\vspace{-0.3cm}
\vspace{-5pt}

\section{Related works}
\vspace{-7pt}

\subsection{Autoregressive speech synthesis}
\vspace{-4pt}

Despite the expressive potential of AR TTS models~\cite{valle,voicecraft,seedtts,lajszczak2024base,du2024cosyvoice}, they have long been plagued by lengthy speech sequences. Compared to text modality, speaking five words often takes one second, where contains fifty speech tokens quantized in 50Hz. Predicting long sequences accumulates the loss from inaccurate decoded tokens, causing instability issues during inference. Generated speech often suffers from bad cases such as unnatural silence, repeated words, and word omissions. 
Hence, robustness is a major concern for current AR TTS models.
A variety of strategies have been proposed in prior work to enhance the robustnes of generation, including alignment-guided sequence reordering or training~\cite{song2024ella,han2024vall}, applying Transducer loss~\cite{du2024vall}, chain-of-thought prompting~\cite{xin2024rall}, etc.

High latency and the substantial semantic gap to the text modality remain major challenges in autoregressive decoding when using fine-grained speech representations. Existing methods to address these issues can be broadly categorized into three approaches:  
(1) Downsampling the representation or utilizing its coarse-grained counterpart;  
(2) Hierarchical generation from coarse to fine representations;  
(3) Employing \textit{multi-token prediction} for chunk-wise AR modeling with carefully designed decoding strategies:

\vspace{-5pt}
\paragraph{Representation downsampling} In recent years, a growing body of work has focused on compressing discrete speech representations to extremely low frame rates~\cite{guo2025recent}. One viable approach is to integrate downsampling modules either directly into the quantization stage~\cite{Siuzdak_SNAC_Multi-Scale_Neural_2024,guo2024speaking,yang2024uniaudio15} of self-supervised representations or within neural speech codec models~\cite{zhang2024speechtokenizer,kyutai2024moshi}. However, such strategies often lead to substantial loss of intra-speech information, coarse-grained speech tokens still struggle to match the performance of fine-grained tokens in capturing acoustic details and other nuanced features.

\vspace{-5pt}
\paragraph{Coarse-to-fine generation} Since fine-grained representations excel in modeling acoustic details, recent studies have explored generating coarse-grained tokens autoregressively, followed by conditioning on these tokens to predict fine-grained representations using either AR or non-AR strategies~\cite{borsos2023audiolm,kharitonov2023speak,wang2024maskgct}.

\vspace{-5pt}
\paragraph{Speech synthesis via chunk-wise multi-token Prediction} Multi-token prediction has been proposed in recent years to accelerate decoding in large language models (LLMs)~\cite{cai2024medusa,gloeckle2024better} and to enable faster convergence during training~\cite{liu2024deepseek}. Recent studies~\cite{vadusa, nguyen2024acceleratingcodecbasedspeechsynthesis, vocalnet} in the speech domain have begun to adopt this approach for efficient autoregressive speech recognition and for fast, high-quality speech synthesis. 

 \vspace{-5pt}
\subsection{Preference policy via reinforcement learning}
 \vspace{-5pt}
Since ChatGPT~\cite{ouyang2022training}, preference optimization policies have attracted significant attention and exploration. 
The Proximal Policy Optimization(PPO)~\cite{PPO} algorithm pioneers in treating human evaluation as supervision of the critic model, which provides environmental rewards when training the fundamental LLM as the actor model. Subsequently, Direct Preference Optimization (DPO)~\cite{DPO} simplified this process by enabling the LLM itself to serve as the reward model. With the recent shift toward the reasoning era, DeepSeek-R1~\cite{guo2025deepseekr1} brought the Group Relative Policy Optimization (GRPO)~\cite{shao2024deepseekmath} algorithm to the forefront of this trend. GRPO leverages inter-group advantages derived from multiple rounds of reasoning by the LLM as the reward signal, effectively stimulating and enhancing the model’s reasoning capabilities. Several studies~\cite{lin2024align,hu2024robust,chen2024enhancing,zhang2024speechalign,yao2025fine} in the speech domain have adapted these preference alignment strategies to advance robust and high-fidelity speech synthesis.

Our method leverages multi-token prediction and reinforcement learning to further optimize the chunk scheduling policy, offering a novel approach to robust, high-quality, and efficient speech synthesis.  

\begin{figure}[t]
\vspace{-5pt}
\centering
\begin{minipage}[t]{\linewidth}
  \begin{minipage}[b]{0.52\linewidth}
    \centering
    \includegraphics[width=\linewidth]{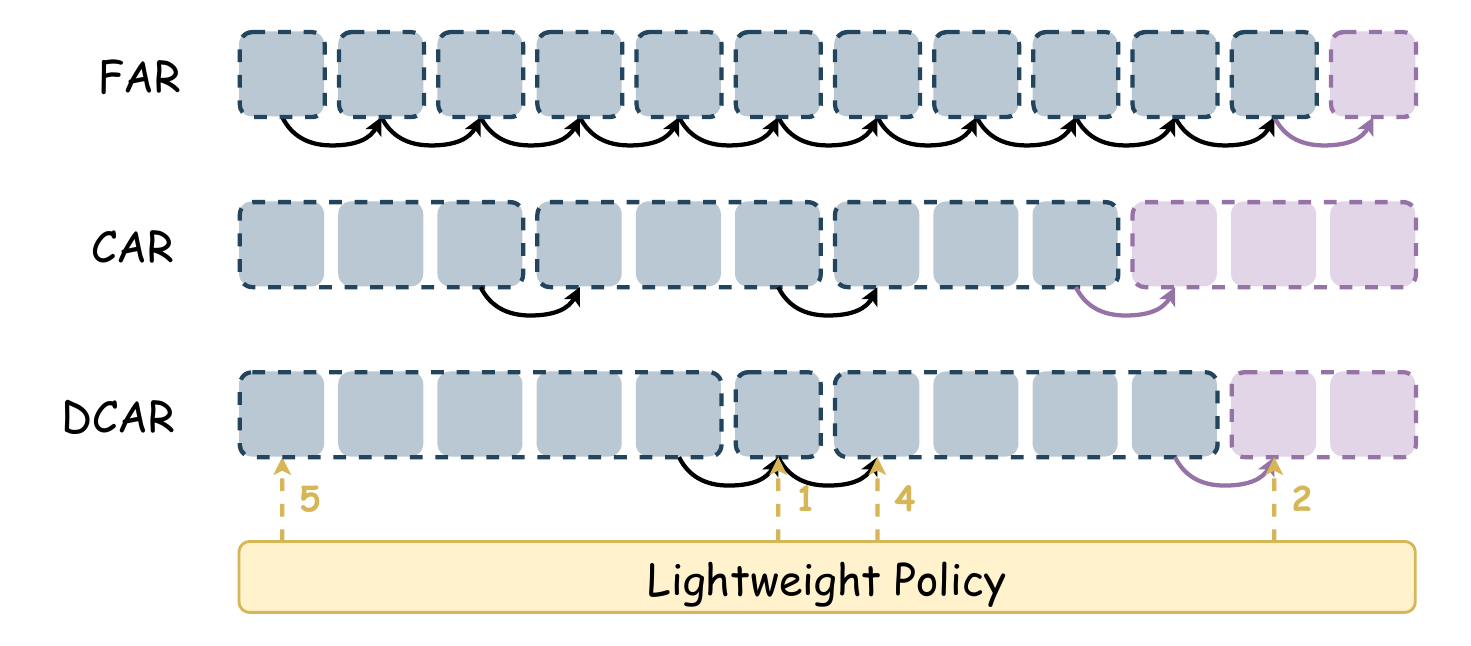}
    \captionof{figure}{Overview of FAR, CAR, and proposed DCAR.}
    \label{fig:chunk2frame}
  \end{minipage}
  \hfill
  \begin{minipage}[b]{0.44\linewidth}
    \centering
    \includegraphics[width=\linewidth]{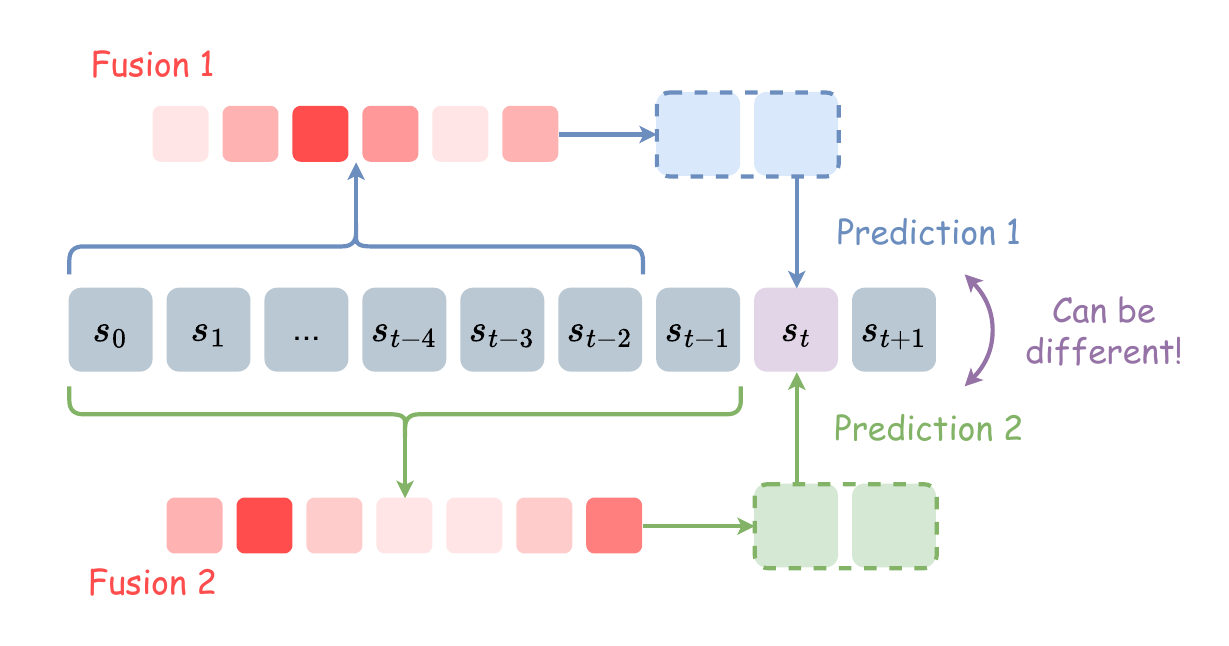}
    \captionof{figure}{Prediction of the same frame varies in different chunks.}
    \label{fig:chunk2frame}
    
  \end{minipage}%
  \hfill
  
  \label{fig:combined-minipage}
\end{minipage}
\end{figure}
\vspace{-10pt}

\section{Revisiting autoregressive TTS with chunk-wise prediction}

In this section, we address a central question: \textbf{\textit{At what scale should each prediction be made in speech synthesis?}}
To explore this, we implement a basic TTS framework based on \textit{multi-token prediction}, where each prediction head is trained with either a declining or uniform objective weighting scheme.
Then we conduct extensive analyses to investigate the reasons behind the superiority of chunk-wise prediction over frame-wise prediction, as well as the limitations that remain in fixed-length chunk-wise prediction.

\subsection{Implementation of chunk-wise autoregressive Text-to-Speech}

\paragraph{Tokenizers}
Two types of speech token are used in building AR TTS models, tokens with a single codebook are considered as a
For semantic tokens, we select the k-means clusters on the last layer of HuBERT-Large~\cite{hsu2021hubert} and $\mathcal{S}^3$-Tokenizer~\cite{du2024cosyvoice}, as unsupervised and supervised semantic tokens respectively.
Specifically, for HuBERT tokens, the k-means model is trained with 2048 centroids on a randomly sampled 73-hour subset of LibriTTS training set.
Previous works \cite{chen2024vall} prove that phoneme and subword perform close as the text type, hence we choose phonemes as input text units for convenience.

\paragraph{Training \textit{CAR} decoder with multiple heads} 
We design additional lightweight heads on the original decoder-only architecture according to ~\cite{cai2024medusa, vadusa}, training them from scratch jointly with bidirectional attention on text tokens and causal attention on speech tokens. Each head is responsible for the prediction of a position in the next chunk sequentially.

The training objective can be described as: 
\begin{equation}
    \mathcal{L}_{\text{CAR}} = -\frac{1}{T}\sum_{t=0}^T\sum_{i=0}^{N'} \gamma^i \log p_\theta(s_{t+i}|s_{<t},x)
\end{equation}

where $s$ refers to the speech token sequence of length $T$, $x$, $\theta$ refer to the text condition and the model parameters, and $\gamma$ is a hyperparameter in the range of $(0, 1]$. Assuming $N$ is the number of the predicting heads, we have \(N' = \min\{N, T-t\}\) in the process. Among a series settings of $\gamma$ values, the equal objective pattern, where $\gamma=1$, performs the best in synthesis robustness. 

\paragraph{Speaker-controllable speech token vocoder}
Semantic tokens like HuBERT clusters and $\mathcal S^3$ Tokenizer tokens do not have a built-in decoder to reconstruct waveforms, hence training an additional vocoder is necessary.
In this work, we use CTX-vec2wav~\cite{du2024unicats} to reconstruct semantic tokens into waveforms, due to its speaker controllability and simple end-to-end procedure. 
Following \cite{guo2024lscodec}, we use WavLM-Large~\cite{chen2022wavlm} to extract speaker representations to best control the speaker identity in inference. 
For $\mathcal {S}^3$-Tokenizer, this structure is faster than the conditional flow matching decoder proposed in \cite{du2024cosyvoice}, which is crucial in the RL training process in later sections.


\subsection{Advantages and limitations of chunk-wise prediction}


We conduct analysis of the foundation of our approach from three key aspects: (1) We revisit the motivation and role of multi-token prediction, and discuss its distinct behavior when applied to the speech modality; (2) Through an analysis of attention weights, we demonstrate how the CAR speech model diverges from the FAR one; (3) We identify the limitations of fixed-length CAR and point out the necessity of adopting a dynamic chunk size. In particular, we argue that the mismatch between training and inference in autoregressive speech generation hinders the learning of a reliable dynamic scheduling policy, thereby motivating our use of reinforcement learning.


\subsubsection{Revisiting chunk-wise token prediction}

In \cite{liu2024deepseek} and \cite{gloeckle2024better}, chunk-wise multi-token prediction is primarily used as an auxiliary task to better exploit training data and accelerate model convergence, while still adopting single-step prediction for best inference. When applied to speed up inference, however, the inherently more challenging nature of cross-token prediction often leads to a degradation in generation quality. MEDUSA \cite{cai2024medusa} attempts to balance inference efficiency and generation quality by promptly verifying additional outputs. A common agreement in the aforementioned studies is that additional prediction tasks are challenging and often yield unreliable results. Nevertheless, due to the stronger local continuity in speech signals (even in form of discrete tokens), we argue that the predicting difficulty of additional tokens is reduced. This diminishes the dominance of the base head in prediction, allowing additional heads to predict comparably well or even better.

\subsubsection{Chunk-to-frame attention pattern}

Chunk-wise prediction models are designed to capture the relationship between an upcoming chunk and the preceding frames. Compared to individual frames, chunks often carry more coherent and semantically informative content, which significantly alters the model's attention patterns. By visualizing the attention weights of our CAR model for a given utterance, we gain insights into how attention is distributed across different positions during chunk prediction.

\begin{figure}[t]
     \centering
     \begin{subfigure}[b]{0.25\textwidth}
         \centering
         \includegraphics[width=\textwidth]{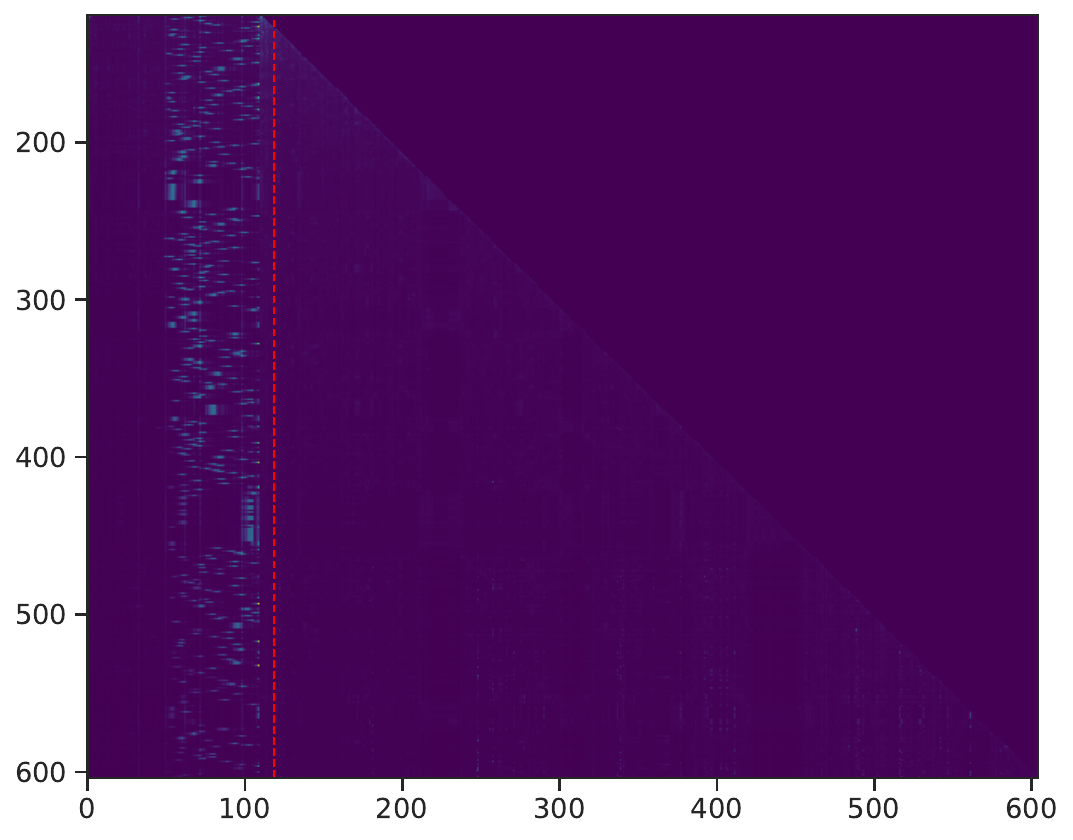}
         \caption{First-layer attention weights from FAR model.}
         \label{fig:mtp0}
     \end{subfigure}
     \hfill
     \begin{subfigure}[b]{0.25\textwidth}
         \centering
         \includegraphics[width=\textwidth]{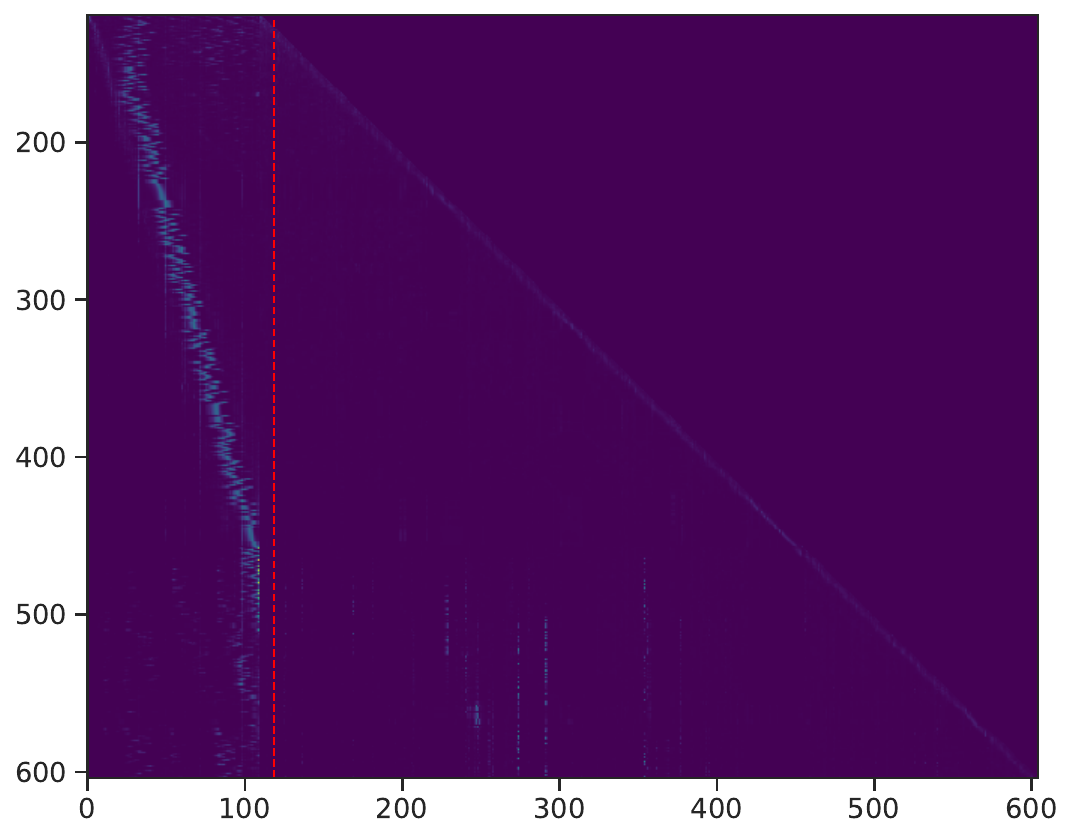}
         \caption{First-layer attention weights from CAR model.}
         \label{fig:mtp6}
     \end{subfigure}
     \hfill
     \begin{subfigure}[b]{0.2\textwidth}
         \centering
         \includegraphics[width=\textwidth]{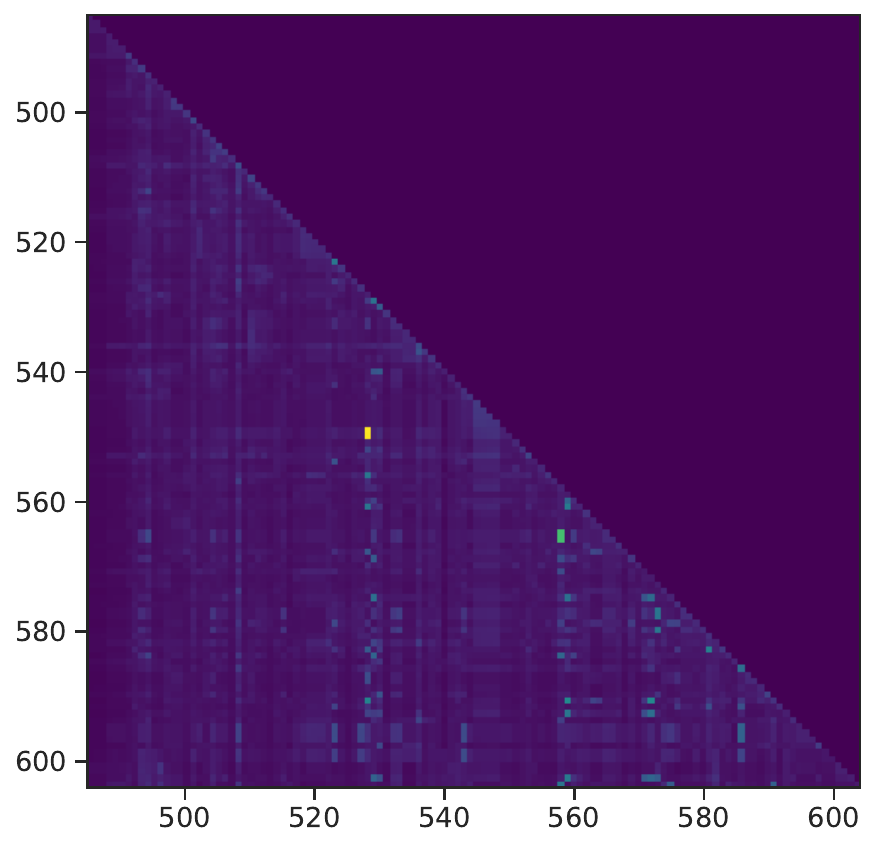}
         \caption{A zoomed-in view of FAR attention.}
         \label{fig:mtp0_local}
     \end{subfigure}
     \hfill
     \begin{subfigure}[b]{0.2\textwidth}
         \centering
         \includegraphics[width=\textwidth]{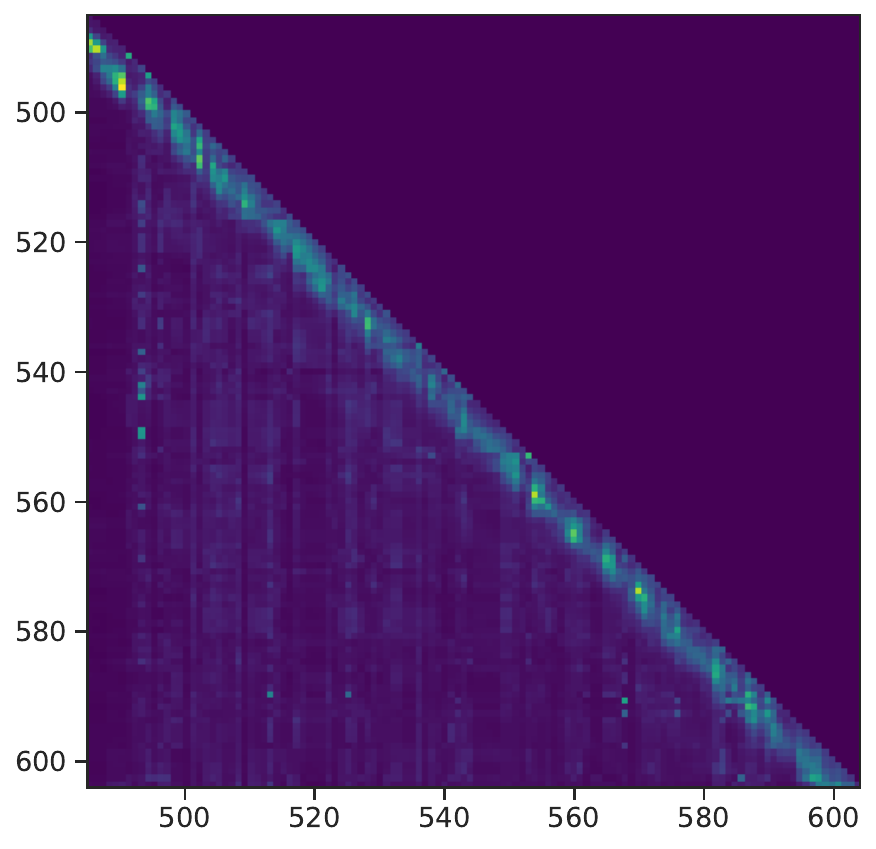}
         \caption{A zoomed-in view of CAR attention.}
         \label{fig:mtp6_local}
     \end{subfigure}
        \caption{Attention visualization. The first 119 tokens are for text, followed by 535 audio tokens.}
        \label{fig:attention_weights}
\end{figure}

Figure~\ref{fig:attention_weights} illustrates the distribution of attention weights in the first layer of both FAR and CAR models. In contrast to pure text models, text-guided speech models exhibit two prominent branches of attention: (1) corresponding text frames, and (2) short-range speech context. Occasionally, the model attends to frames more than several dozen steps away, which we hypothesize is due to either speaker-specific characteristics or genuinely similar acoustic patterns.

Comparison results in Figure~\ref{fig:attention_weights} reveals a notable distinction between the two models. The CAR model demonstrates a clearer alignment between speech and text in earliest layers, as well as a more stable focus on short-range speech context. In contrast, the FAR model shows signs of confusion in these aspects. This discrepancy can be reasonably attributed to the chunk-level representation being inherently more informative and easier to interpret than individual frames.

It is also worth noting that the attention distribution over short-range speech context is highly uneven across positions. Figure~\ref{fig:attention_weights}\textnormal{(c)(d)} zoom in on the local attention structure near the diagonal, showing that attention within a chunk is often diffusely allocated to the preceding few frames. This pattern is consistent with the nature of speech, where certain key frames in the sequence largely determine the trajectory of speech content.


\subsubsection{Frame-to-frame attention preference}


For a certain frame, the predictions from different chunks can be different. In Figure~\ref{fig:chunk2frame}, as an example with a chunk size of 2, if the prediction of $s_t$ from $p_{\theta}(s_t, s_{t+1} | s_{<t})$ is weaker than that from $p_{\theta}(s_{t-1}, s_t | s_{<t-1})$, we should not simply attribute this to the presence of the additional $s_{t-1}$ in the condition or the inclusion of an extra prediction target $s_{t+1}$. Instead, we consider it as evidence that, given the model’s limited capacity, the information fused for chunk $(s_t, s_{t+1})$ is more suitable for predicting $s_t$ than that for chunk $(s_{t-1}, s_t)$. In other words, each frame inherently has a preference over different chunk-to-frame attention patterns.



\begin{figure}[t]
  \centering
  \includegraphics[width=\linewidth]{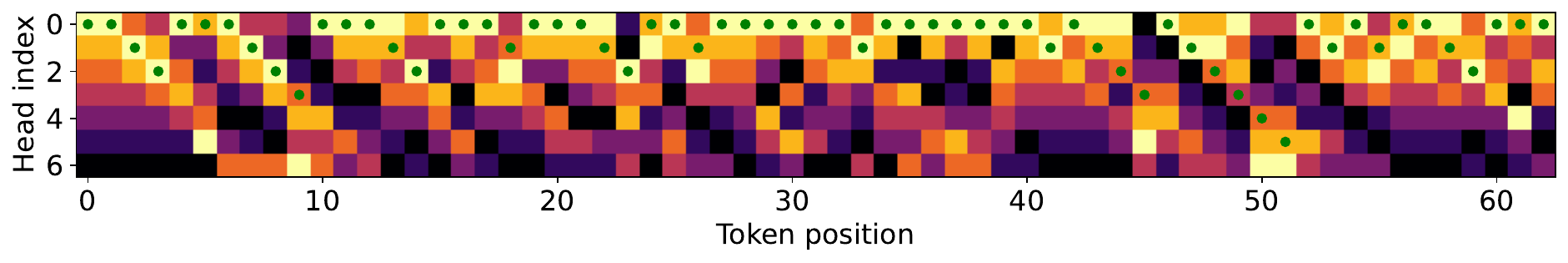}

  \caption{Prediction loss ranking across different heads (teacher forcing mode). Brighter colors indicate better prediction performance for each position. The green dots illustrate a dynamic prediction route that selects prediction as well as possible.}
  \label{fig:posranks}
  \vspace{-10pt}
\end{figure}

Figure~\ref{fig:posranks} illustrates the inference behavior of a CAR model with 6 additional heads on a GT sequence. We slide the 7 prediction heads of CAR with a stride of one frame, so predictions within each chunk are aligned diagonally. In teacher forcing mode, we compute cross-entropy loss at each position and sort them within each column, which correspond to the predictions for the same position made when the base head is at different positions ahead. A lower loss (represented by brighter colors in the figure) indicates that the model is more confident in predicting the GT token.

We observe that the most confident prediction for a given position often does not come from the base head. Experiments on the \textit{dev-all} subset of LibriTTS show that the base head makes the best prediction in only 60.07\% of the cases while producing 9.27\% worst predictions.

When considering minimizing the total prediction loss under teacher forcing, the problem becomes one of dynamic programming. A heuristic principle is to follow the path through brighter cells as many as possible, as shown in Figure~\ref{fig:posranks}, with each diagonal representing a single prediction step. 

However, such principle is merely heuristic and is difficult to leverage for training an effective scheduling policy. This is partly because it fail to account for the varying importance of different positions. More critically, the model's behavior during actual inference differs from that under teacher forcing: \textbf{In AR generation, divergences from the ground-truth sequence may lead to accumulated errors, but can also result in alternative, satisfactory outputs.} Despite this, single-step FAR or fixed-length CAR evidently leaves more room for optimization.

This particular nature in AR poses challenges to learning the policy under teacher forcing. We argue that a reasonable scheduling policy need to consider the overall trajectory across multiple future frames, and thus differs from speculative sampling. Therefore, we aim to obtain feedback on the scheduling policy directly from the generation results, which motivates our use of reinforcement learning. We believe that reinforcement learning can address such difficulties, and that feedback derived directly from audio metrics is unconstrained by the ground-truth data and brings potential for further improving speech synthesis performance.



\section{Dynamic chunk-wise prediction via lightweight policy}

\vspace{-5pt}
\subsection{Preliminaries}
\vspace{-5pt}

GRPO is a varient of PPO, which foregoes the critic model and use group-related advantage values to instead. It can directly utilize rule-based rewards, performing as an efficient and effective RL algorithm. In the training process, the updating policy is $\pi_{\theta}$, reference policy is $\pi_{\theta_{\text{ref}}}$ and the old policy is $\pi_{\theta_{\text{old}}}$. Given a question $q$ from set $Q$, a group of $G$ candidate outputs $\{o_1, o_2, \cdots, o_G\}$ is generated using the $\pi_{\theta_{\text{old}}}$. The advantage $\hat{A}_{i,t}$ will be constructed utilizing rewards, and the training objective is:

\begin{align}
\mathcal{J}_{\text{GRPO}}(\theta) = 
\mathbb{E}_{q \sim P(Q),\, \{o_i\}_{i=1}^{G} \sim \pi_{\theta_{\text{old}}}(\mathcal{O}|q)} 
\notag 
\Bigg[
\frac{1}{G} \sum_{i=1}^{G} \frac{1}{|o_i|} \sum_{t=1}^{|o_i|} 
\min \Bigg(
    \frac{\pi_\theta(o_{i,t} | q, o_{i,<t})}{\pi_{\theta_{\text{old}}}(o_{i,t} | q, o_{i,<t})} \, \hat{A}_{i,t}, \notag \\
\quad \text{clip} \left(
        \frac{\pi_\theta(o_{i,t} | q, o_{i,<t})}{\pi_{\theta_{\text{old}}}(o_{i,t} | q, o_{i,<t})}, 
        1 - \varepsilon, 1 + \varepsilon
    \right) \hat{A}_{i,t}
\Bigg) \Bigg] 
- \beta \, \mathbb{D}_{\mathrm{KL}}[\pi_\theta \| \pi_{\text{ref}}]
\end{align}

where $\varepsilon$ and $\beta$ are hyperparameters, and $\mathbb{D}_{\mathrm{KL}}[\pi_\theta \| \pi_{\text{ref}}]$ can be described as:

\vspace{-5pt}

\begin{equation}
\mathbb{D}_{\mathrm{KL}} \left[ \pi_\theta \| \pi_{\text{ref}} \right] = 
\frac{\pi_{\text{ref}}(o_{i,t} | q, o_{i,<t})}{\pi_\theta(o_{i,t} | q, o_{i,<t})}
- \log \frac{\pi_{\text{ref}}(o_{i,t} | q, o_{i,<t})}{\pi_\theta(o_{i,t} | q, o_{i,<t})}
- 1
\end{equation}

\subsection{Dynamic chunk-wise policy optimization}
\label{dcpo}
We design our task-specific RL algorithm, \textit{Dynamic Chunk-wise Policy Optimization} (DCPO), for training a lightweight dynamic chunk-wise schedule policy as an adaptive modification of GRPO with some specific training strategies. We select a tiny subset of samples from LibriTTS train set, which contains 980 utterances from 20 speakers with durations at least 6 seconds. The first 3 seconds will be utilized as an audio prompt.

\vspace{-7pt}

\paragraph{Training objective} The architecture of the schedule module is a linear predicting head followed by a causal transformer layer, which serves the sequence of the historical hidden states of predicting speech sequences as the input. Hence, our training objective should take the parameters $\phi$ of the base model into consideration:
\begin{align}
\label{eq4}
\mathcal{J}_{\text{DCPO}}(\theta) = 
\mathbb{E}_{q \sim P(Q),\, \{a_i\}_{i=1}^{G} \sim \pi_{\theta_{\text{old}}}(\mathcal{A}|q)} 
\notag 
\Bigg[
\frac{1}{G} \sum_{i=1}^{G} \frac{1}{|a_i|} \sum_{t=1}^{|a_i|} 
\min \Bigg(
    \frac{\pi_\theta(a_{i,t} | q, s(a_{i,<t}, \phi))}{\pi_{\theta_{\text{old}}}(a_{i,t} | q, s(a_{i,<t}, \phi))} \, \hat{A}_{i,t}, \notag \\
\quad \text{clip} \left(
        \frac{\pi_\theta(a_{i,t} | q, s(a_{i,<t}, \phi))}{\pi_{\theta_{\text{old}}}(a_{i,t} | q, s(a_{i,<t}, \phi))}, 
        1 - \varepsilon, 1 + \varepsilon
    \right) \hat{A}_{i,t}
\Bigg) \Bigg] 
- \beta \, \mathbb{D}_{\mathrm{KL}}[\pi_\theta \| \pi_{\text{ref}}]
\end{align}
Here, the output set $\mathcal{A}$ comprises lightweight policy action sequences, and the function $s$ computes the base model's hidden states. Specifically, $s$ leverages previously determined policy actions to schedule chunk-wise decoding steps and subsequently samples tokens as base model input, then get hidden states accordingly. In the forward pass, we introduce a mask on inner-chunk positions, ensuring the policy only learns from positions that really schedule chunk size in the sampling. In practice, we use a warm-up strategy for $\beta$, initializing it at 0 and incrementally increasing it over training epochs. Additionally, the advantage $\hat{A}_{i,t}$ is defined as $\frac{r_i - \text{mean}(\mathbf{r})}{\text{std}(\mathbf{r})}$ following GRPO’s outcome-supervision methodology, with the reward vector $\mathbf{r}$ detailed below.

\vspace{-7pt}

\paragraph{Chase-then-exceed strategy} To motivate the random-initialized policy capability for robust synthesis, we profile a range $\mathcal{C}$ of chunk-wise schedule actions which outperforms in \textit{CAR} speech synthesis and set half of the sample group \textit{CAR}-scheduled using fixed chunk length as the values in the range. This softly constrains the action searching space, enabling the policy first learning from \textit{CAR} experience and then outperform \textit{CAR}.

\vspace{-7pt}

\paragraph{Reward function via action guidance} During training, we sample batches from the tiny dataset, decoding each of them $G$ times as groups. After utilizing an efficient automatic speech recognition model to transcribe the generated samples and the ground truth speech and then compute their WER metrics for outcome rewards. We give a value-relative negative process rewards for the out-range actions as an additional constrain. The calculation function for the total rewards $r$ is formulated as:

\begin{equation}
\label{eq5}
    r = \max \{1-\ln(\text{WER}_{\text{gen}}-\text{WER}_{\text{gt}}+1), \epsilon\} - \lambda \sum_{i \notin {\mathcal{C}}}\frac{i}{T}
\end{equation}

 \vspace{-10pt}
 
where $\text{WER}_{\text{gen}}$ and $\text{WER}_{\text{gt}}$ refer to the WER metrics for generated and ground truth speech and $T$ refers to the generated speech sequences length. The value of $\epsilon$ is $-10$ for avoiding reward explosion and $\lambda$ is usually set as 0.1, enabling the absolute value of total negative rewards an order of magnitude less than the WER supervision reward.

\vspace{-5pt}
\begin{table*}[t]
  \centering
  \small
  \begin{tabular}{l|cccccc}
    \toprule
    \multirow{2}{*}{\textbf{Method}} 
      & \multirow{2}{*}{\textbf{Avg.\ Token}}
      & \multicolumn{3}{c}{\textbf{Quality}} 
      & \multicolumn{2}{c}{\textbf{Efficiency}} \\
      & 
      & WER$\downarrow$ & SECS$\uparrow$ & UTMOS$\uparrow$ 
      & RTF$\downarrow$ & Speedup$\uparrow$ \\
    \midrule

    \multicolumn{7}{c}{\textit{\textbf{HuBERT-kmeans2048-50Hz}}} \\
    \midrule
    FAR                     & 1    & 9.99 & 0.885 & 4.12$_{\pm0.030}$ & 0.320 & 1$\times$ \\
    \midrule
    \multirow{3}{*}{CAR -- WER top3}
                            & 2   & 3.04 & \textbf{0.886} & 4.14$_{\pm0.028}$ & 0.171 & 1.89$\times$ \\
                            & 3   & 2.99 & 0.885 & 4.14$_{\pm0.028}$ & 0.121 & 2.83$\times$ \\
                            & 4   & 3.09 & 0.885 & 4.13$_{\pm0.027}$ & \textbf{0.094} & \textbf{3.60}$\times$ \\
    \midrule
    \textbf{\MethodName~-- {[}2,3{]}}   & 2.78   & 2.82 & 0.885 & \textbf{4.14$_{\pm0.027}$} & 0.137 & 2.39$\times$ \\
    \textbf{\MethodName~-- {[}2,3,4{]}} & 3.08   & \textbf{2.77} & 0.885 & 4.14$_{\pm0.028}$ & 0.128 & 2.61$\times$ \\
    \midrule

    \multicolumn{7}{c}{\textit{\textbf{$\mathcal{S}^3$-Tokenizer-50Hz}}} \\
    \midrule
    FAR                     & 1    & 2.31 & 0.884 & 4.26$_{\pm0.022}$ & 0.335 & 1$\times$ \\
    \midrule 
    \multirow{3}{*}{CAR -- WER top3}
                            & 2   & 2.02 & \textbf{0.885} & 4.27$_{\pm0.019}$ & 0.171 & 2.04$\times$ \\
                            & 3   & 2.09 & 0.884 & 4.27$_{\pm0.017}$ & 0.121 & 2.94$\times$ \\
                            & 4   & 2.06 & 0.883 & 4.27$_{\pm0.018}$ & \textbf{0.096} & \textbf{3.76}$\times$ \\
    \midrule
    \textbf{\MethodName~-- {[}2,3{]}}   & 2.41 & \textbf{1.84} & \textbf{0.885} & \textbf{4.29$_{\pm0.017}$} & 0.158 & 2.18$\times$ \\
    \textbf{\MethodName~-- {[}2,3,4{]}} & 3.27   & 1.92 & 0.884 & 4.26$_{\pm0.018}$ & 0.122 & 2.89$\times$ \\
    \bottomrule
  \end{tabular}
  \caption{Performance of FAR, CAR, DCAR on speech synthesis robustness, quality, and speed. ``Avg. Token'' refers to the average number of speech token simultaneously generated in each step.}
  \label{tab:table-main}
  \vspace{-5pt}
\end{table*}

\section{Experiments}
\vspace{-5pt}

\subsection{Experimental settings}
\vspace{-3pt}
\paragraph{Datasets}
We conduct our main experiments on LibriTTS~\cite{libritts} training set with 585 hours, and LibriHeavy with 50k hours is utilized for data scaling experiments(Appendix \ref{E2}). For evaluation, we test TTS performance on \textit{UniCATS testset-B}~\cite{du2024unicats}. This set contains 500 utterances from 37 unseen speakers in the LibriTTS test-clean subset. 
Each speaker is associated with a speech prompt of approximately 3 seconds. The policy training set is described in section ~\ref{dcpo}.

\vspace{-7pt}

\paragraph{Architecture configurations}
The fundamental decoder-only TTS architecture is constructed with 12 Transformer layers, each layer with 8 attention heads, 1024 hidden dimensions, and 4096 feedforward dimensions. 
The base predicting head is a linear layer, and the additional heads consist of 4 residual blocks (a linear layer plus a SiLU~\cite{elfwing2018sigmoid} activation function) and a linear prediction layer. For fair comparison, we conduct our main experiments on the setting of 6 addition heads.
In training, we compute WER using NeMo ASR\footnote{\scriptsize\texttt{https://huggingface.co/nvidia/stt\_en\_fastconformer\_transducer\_large}}.

\vspace{-7pt}

\paragraph{Baselines} We select three methods as baselines of DCAR: (1) Frame-level AR TTS with next-token prediction; (2) Chunk-wise AR TTS selecting fixed-number tokens as the next chunk; (3) VADUSA~\cite{vadusa}, a speculative decoding strategy selecting draft tokens through the verification based on the predicting head, which also performs dynamic effect in decoding steps. Except for whether having additional heads, all the baselines are in the same architecture.

\vspace{-7pt}

\paragraph{Evaluation metrics} To evaluate speech synthesis performance comprehensively, we utilize metrics for various aspects: \textit{Word-error-rate} (WER) and UTMOS~\cite{saeki22c_interspeech} for synthesis intelligibility and quality; \textit{speaker encoder cosine similarity} (SECS) for zero-shot TTS speaker similarity; and \textit{real time factor} (RTF) for synthesis speed. 
We measure WER by Whisper-Large-V3\footnote{Note that this is different with the ASR tool NeMo in guidance range profiling and on-policy training, to avoid potential overfitting.}~\cite{whisper}, and SECS by Resemblyzer\footnote{\scriptsize\texttt{https://github.com/resemble-ai/Resemblyzer}}.
We also measure the \textit{\MethodName} speedup effect compared to the baselines.
\vspace{-7pt}

\paragraph{Devices}
Our main experiments are conducted on NVIDIA RTX 4090 24GB GPUs, and data-scaled training is conducted on NVIDIA A800 80GB GPUs. 


\vspace{-5pt}
\subsection{Performance of different AR generation paradigms}
\vspace{-3pt}

In Table~\ref{tab:table-main}, we observe that the FAR-based generation method underperforms in terms of robustness across both token types. It also fails to achieve optimal results in zero-shot speaker similarity (SECS) and in naturalness as measured by UTMOS. Furthermore, this method incurs significant inference latency. The table also reports the top-three most robust results for the CAR-based synthe method. We find that simply training the speech synthesis model using the CAR architecture leads to substantial robustness improvements. Additionally, its inference speed increases proportionally with the number of predetermined decoding tokens.
DCAR achieves the lowest WER in robustness assessments. Importantly, it maintains acoustic quality and zero-shot performance while also providing considerable acceleration. Notably, when the average sampling steps are comparable, DCAR's speedup is slightly diminished due to the overhead introduced by the policy model. 

\vspace{-5pt}
\subsection{Comparison between DCAR and speculative decoding}
\vspace{-3pt}

While speculative decoding methods like \textit{VADUSA} aim to accelerate inference, DCAR leverages WER as the primary supervision signal to enhance the robustness of speech synthesis. Despite differing in both objectives and underlying mechanisms, the two approaches converge on a similarly dynamic decoding strategy, ultimately achieving high-quality and efficient speech generation. We ensure architectural consistency in the CAR TTS models and demonstrate the generation quality and speed achieved by both methods using the shared HuBERT tokens in Table~\ref{vs. vadusa}. \textit{VADUSA} proposes tolerant verification on multiple sampling results of the base head, here we set the tolerance value as 2, 3.

\vspace{-5pt}
\subsection{Discussion on the frame-rate factor}
We compare and integrate the CAR framework with low-frame-rate tokenization approaches, specifically utilizing both 50Hz and 25Hz variants of the $\mathcal{S}^3$-Tokenizer. Given that 25Hz tokens cover a temporal receptive field twice as large as that of 50Hz tokens, we adapt the CAR model by reducing the number of prediction heads to three—half the number used in the 50Hz configuration. To assess the effectiveness of CAR, we provide a visualization in Figure~\ref{fig:temporal}, with the x-axis representing the temporal dimension. As shown, the generation performance of CAR degrades with increasing chunk sizes in the 25Hz setting. Nevertheless, over half of these configurations still outperform the FAR baseline, while offering up to 2 times theoretical acceleration in synthesis. On the other hand, CAR synthesis using 50Hz tokens consistently outperforms the 25Hz CAR counterparts when controlling for the temporal field of view. Furthermore, we report the performance of the 50Hz DCAR model, which achieves the highest robustness across all evaluated settings.

\vspace{-5pt}
\subsection{Ablation study}

\vspace{-5pt}

\paragraph{Variation of policy action guidance range}

Table~\ref{DCPO action guidance range} demonstrates the impact of changing the action guidance range of DCPO algorithm. Wider range of the profiled CAR chunk size indicates to faster decoding ability of the policy, with declining of robust synthesis capability.
\vspace{-10pt}

\paragraph{Method Decomposition} We decompose the DCPO strategies,  chase-then-exceed strategy and outrange negative rewards, as demonstrated in Table~\ref{decomposition}, to reflect the effectiveness of designed strategies. ``Random'' policy in the table directly means select the chunk size at each decoding step randomly from the followed range.



\begin{table}[t]
\centering
\small
\begin{minipage}[t]{0.48\linewidth}
    \centering
    \begin{tabular}{l|c}
    \toprule
    \textbf{Policy} & \textbf{WER(\%)}$\downarrow$ \\
    \midrule
    DCAR -- {[}2, 3{]} & 2.82 \\
    \quad-- w/o. \textit{chase-then-exceed} & 3.17 \\
    \quad-- w/o. outrange neg. rewards & 3.09 \\
    \bottomrule
    \vspace{2pt}
    \end{tabular}
    
    \captionof{table}{Strategy decomposition results in WER.}
    \label{decomposition}
    \centering
    \includegraphics[width=\linewidth]{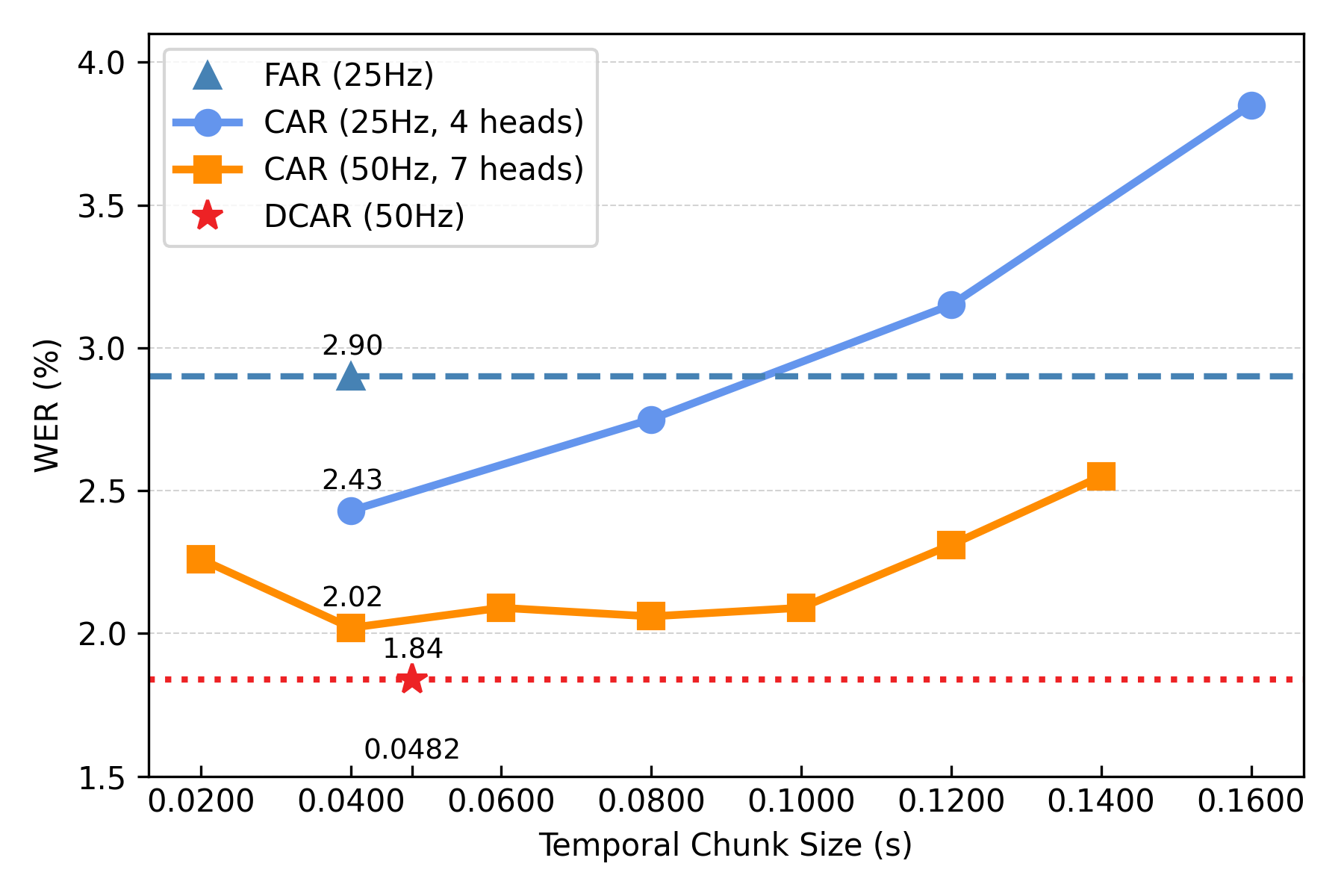}
    \captionof{figure}{WER performance with different frame-rate and strategies.}
    \label{fig:temporal}
\end{minipage}\hfill
\hfill
\begin{minipage}[t]{0.48\linewidth}
    \centering
    \begin{tabular}{lcc}
    \toprule
    \textbf{Method} & \textbf{WER(\%)}$\downarrow$ & \textbf{Speedup}$\uparrow$ \\
    \midrule
    VADUSA--$\tau$=2 & 3.86 & 3.03$\times$ \\
    VADUSA--$\tau$=3 & 4.17 & \textbf{3.12}$\times$ \\
    DCAR--{[}2,3{]} & 2.82 & 2.39$\times$ \\
    DCAR--{[}2,3,4{]} & \textbf{2.77} & 2.61$\times$ \\
    \bottomrule
    \vspace{2pt}
    \end{tabular}
    \captionof{table}{Comparison with VADUSA, $\tau$ refers to the value of tolerance.}
    \label{vs. vadusa}
    \begin{tabular}{l|cc}
    \toprule
    \textbf{DCPO range} & \textbf{WER(\%)}$\downarrow$ & \textbf{RTF}$\downarrow$ \\
    \midrule
    FAR-baseline & 9.99 & 0.320 \\
    {[}2, 3{]} & 2.82 & 0.137 \\
    {[}3, 4{]} & 2.84 & 0.126 \\
    {[}2, 3, 4{]} & \textbf{2.77} & 0.128 \\
    {[}2, 3, 4, 5{]} & 3.04 & \textbf{0.115} \\
    Totally free & 3.30 & 0.118 \\
    \bottomrule
    \label{DCPO action guidance range}
    \end{tabular}
    \vspace{2pt}
    \caption{DCPO action guidance range effects on DCAR synthesis robustness and inference speed.}

\end{minipage}
\vspace{-5pt} 
\end{table}

\vspace{-5pt}
\section{Conclusion}

In conclusion, this work addresses a critical challenge in speech synthesis: the limitations of frame-level autoregressive (FAR) models, which often suffer from instability and high inference latency when operating on fine-grained speech tokens. To overcome these issues, we analyze the effectiveness of chunk-wise autoregressive (CAR) synthesis as an alternative framework. By leveraging its architectural properties, CAR exhibits significant advantages over FAR in both robustness and efficiency. Building on this foundation, we introduce DCAR, a dynamically scheduled variant of CAR that employs reinforcement learning to train a lightweight policy network. This policy adaptively determines optimal chunk sizes at each decoding step, further enhancing synthesis quality and efficiency. Limitations and broader impacts are further discussed in the appendix provided in the supplementary materials.

\newpage
\bibliographystyle{IEEEtran}
\bibliography{neurips_2025}
\newpage





\appendix

\newpage
\section*{\centering Appendix}
\startcontents[sections]
\printcontents[sections]{l}{1}{\setcounter{tocdepth}{2}}
\newpage

\section{Limitations and broader impacts}
\label{limitations and broader impact}

\paragraph{Limitations} Beyond demonstrating the potential of our method for speech synthesis, this paper acknowledges its limitations.

First, DCAR introduces a new paradigm for speech synthesis that substantially reduces the instability seen in FAR and CAR. However, it does not eliminate all failure modes—every method for AR speech synthesis robustness, to some degree, shares this limitation. Combining DCAR with complementary techniques from prior work may offer a promising path toward fully addressing these residual “bad‐case” scenarios.

Another limitation of DCAR is its focus on synthesis robustness at the expense of flexible decoding‐speed control.  In practice, one must re‐tune the DCPO guidance range to bias the policy toward taking more actions—trading off speed against stability—which entails a lightweight retraining step.  By contrast, truly training‐free acceleration methods let users adjust decoding configurations on the fly to achieve faster inference without any additional model updates. However, owing to the lightweight property of DCPO policy, its additional training overhead is minimal.

\paragraph{Broader impacts} DCAR is a versatile speech‐synthesis approach that can be integrated into any autoregressive architecture. Beyond the Text‐to‐Speech experiments presented in this work, it is equally applicable to end‐to‐end spoken‐dialogue systems. Such systems typically employ discrete speech tokens within an AR framework and demand both real‐time responsiveness and synthesis robustness—capabilities that align naturally with the strengths of DCAR. Moreover, our analysis of the chunk-wise AR synthesis (CAR) paradigm offers a fresh perspective on autoregressive speech synthesis, highlighting its potential not only for discrete speech representation generation but also for continuous-space AR synthesis—and inviting deeper exploration in this direction.
\section{Details of DCPO algorithm}
\label{DCPO algorithm details}

\subsection{Action guidance profile}
We leverage the NeMo ASR model for rapid profiling, selecting an action guidance set $\mathcal{C}$ from CAR chunk sizes that achieve top-$k$ performance on the WER metric (where $k$ is a hyperparameter denoting the size of the set, i.e., $|\mathcal{C}|$).

\subsection{Description}

To provide a detailed description of the DCPO algorithm, we present the procedure in Algorithm~\ref{alg:dcpo}.

\begin{algorithm}[htbp]
\caption{Dynamic Chunk-wise Policy Optimization}
\label{alg:dcpo}
\begin{algorithmic}[1]
\Require the policy model $\pi_\theta$; prompts $Q$; action guidance range $\mathcal{C}$; CAR TTS model $\mathcal{S}_\phi$; ASR model $\mathcal{T}$; Vocoder $\mathcal{V}$; hyperparameters $\varepsilon$, $\lambda$.
\State Initialize $\pi_\theta \leftarrow \pi_{\theta_\text{init}}$.
\For {$n$ = $1,...,N$}
    \State Update reference model $\pi_\text{ref} \leftarrow \pi_\theta$.
    \State Set hyperparameter: $\beta \leftarrow 0.1\times (n-1)$.
    \For{$b$ = $1,...,B$}
        \State Sample a batch $Q_b$ from $Q$.
        \State Update the old policy model $\pi_{\theta_\text{old}}$.
        \State Set $|\mathcal{C}|$ actions by traveling the guidance range $\mathcal{C}$.
        \State Set another $|\mathcal{C}|$ actions by $\pi_{\theta_\text{old}}(\cdot| \mathcal{S}_{\phi}(q))$ for each prompts $q \in Q_b$. 
        \State Merge the questions to get the action group $\{a_i\}_{i=1}^{G}$, where $G=2|\mathcal{C}|$.
        \State Compute the rewards $\{r_i\}_i^{G}$ for $a_i$ by scheduling $\mathcal{S}_\phi(q)$ and then running $\mathcal{V}$, $\mathcal{T}$ (Equation~\ref{eq5}).
        \State Compute $\hat{A}_{i,t}$ for the $t$-th action of $a_i$ through $\hat{A}_{i,t}=\frac{r_i - \text{mean}(\mathbf{r})}{\text{std}(\mathbf{r})}$.
        \For{$i$ = $1, ..., I$}
            \State Update the policy model $\pi_\theta$ by maximizing the DCPO objective (Equation~\ref{eq4}).
        \EndFor
    \EndFor
\EndFor
    
\end{algorithmic}

\end{algorithm}

\section{Training details}
\label{training details}
\subsection{CAR TTS configurations}
We utilize 12 layers of causal transformer layer for TTS decoder, each of which consists of 1024 hidden dimensions and 4096 feedforward dimensions. The model has 512 text token embeddings and 2049 speech token embeddings for HuBERT tokens or 4097 speech token embeddings for $\mathcal{S}^3$-Tokenizer tokens(codebook + [EOS] token). We train our model on 8 NVIDIA RTX4090 GPUs for 20 epochs. LibriTTS training subset utterances are filtered to durations between 3s and 20s. We use a per-GPU batch size of 2 and accumulate gradients over 2 steps (effective batch size = 32). Training is performed with ScaledAdam~\cite{yao2023zipformer} optimizer(initial leaning rates = 0.01, $\beta$=(0.9, 0.95) and gradient clipping scale = 2.0) and  Eden scheduler~\cite{yao2023zipformer} (200 warm-up steps)\footnote{The implementation follows: https://github.com/lifeiteng/vall-e}.

We augment the decoder with additional heads for parallel CAR prediction branches, each comprising 4 stacked residual blocks followed by a bias‐free linear projection into the output space with dimension of speech token number. Each residual block applies a $1024\times1024$ linear transform and SiLU activation. The critiron for all the heads is the cross entropy loss. The vocoder is trained for 1 million steps, whose hyper-parameters follow CTX-vec2wav~\cite{du2024unicats}.

\subsection{DCPO training configurations}

The DCPO lightweight policy is a causal transformer layer with 1024 hidden size followed by a $1024\times C$ linear layer predicting $C$ actions. We utilize Adam optimizer with $2e-6$ learning rates and the clip epsilon $\varepsilon$ is set as 0.2. For $\beta$, the weight of KL loss, we schedule it as: $0.1\times(\text{current\_epoch\_number}-1)$, and for the total number of training epochs, we set it as 3. The on-policy training approximately take 8 hours on one NVIDIA RTX4090 GPU.

\section{Evaluation Metrics}
\label{metrics}

\paragraph{WER}
Word Error Rate (WER) is the a standard metric for evaluating intelligibility of generated speech.  It is defined as
\[
\mathrm{WER} = \frac{S + D + I}{N},
\]

where $S$, $D$, and $I$ denote the numbers of substitutions, deletions, and insertions, respectively, and $N$ is the total number of words in the reference transcript.  A lower WER indicates better intelligibility.

\paragraph{SECS} The SCES (Speaker Embedding Cosine Similarity) metric quantifies how closely the synthesized voice matches the target speaker’s voice. It is computed by encoding both the generated and reference utterances with a fixed speaker‐embedding model and taking the cosine similarity between their embeddings—values closer to 1 indicate stronger speaker resemblance.

\paragraph{UTMOS} UTMOS (UTokyo‐SaruLab MOS)~\cite{saeki22c_interspeech} is an objective mean‐opinion‐score predictor that uses an ensemble of trained neural models to estimate human listening ratings on a 1–5 scale. It provides a reliable, reproducible proxy for subjective audio quality assessment without requiring labor‐intensive listening tests. We also report UTMOS’s standard error alongside its mean score to ensure robustness of our evaluations.

\paragraph{RTF} Real‐Time Factor (RTF) measures synthesis speed by dividing the total inference time by the duration of the generated audio.

\section{Expansion experiments}

\subsection{Overview TTS performance}
\begin{minipage}[t]{0.48\linewidth}
    
    \includegraphics[width=\linewidth]{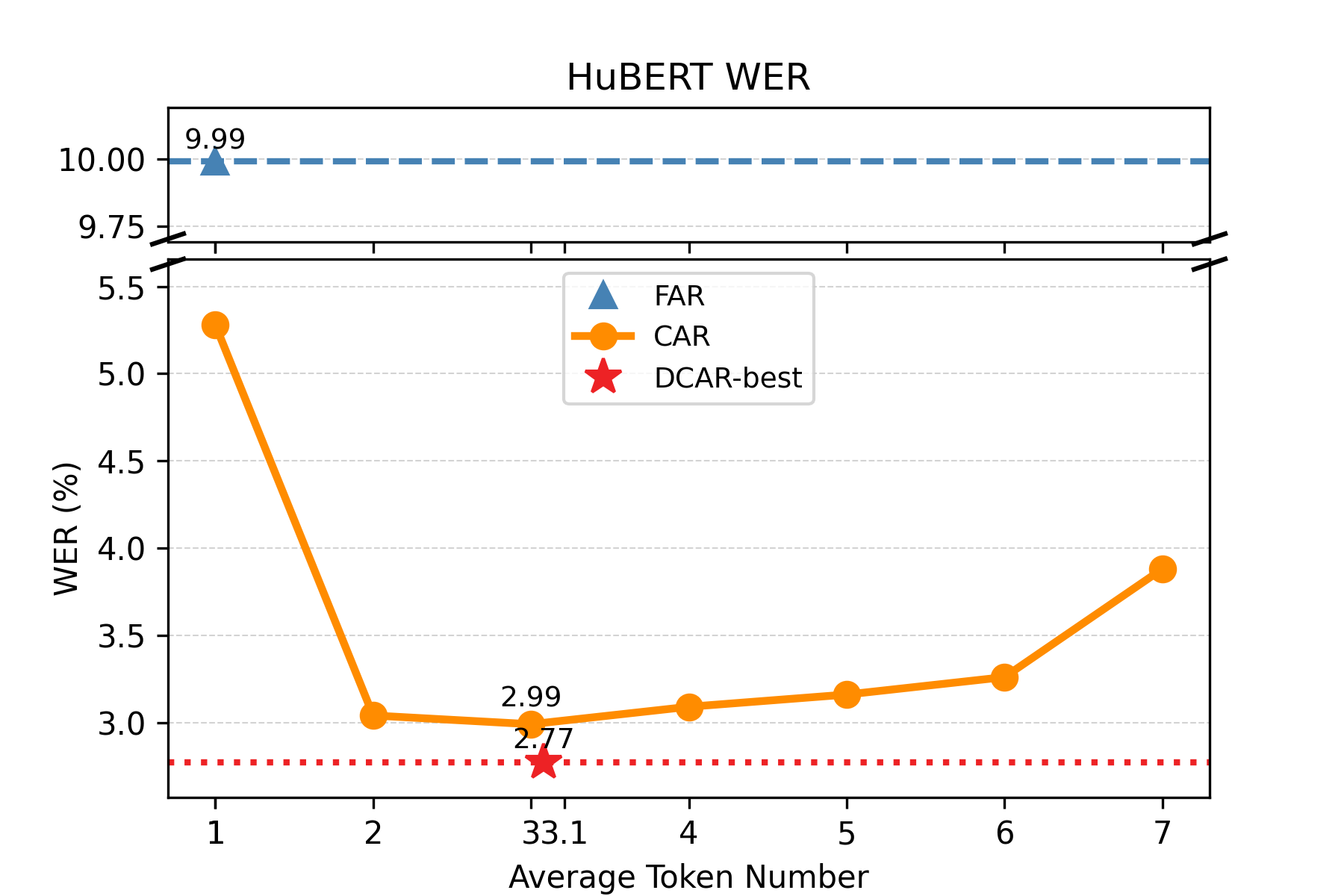}
    \captionof{figure}{}
    \label{fig:hubert}
\end{minipage}
\begin{minipage}[t]{0.48\linewidth}
    \includegraphics[width=\linewidth]{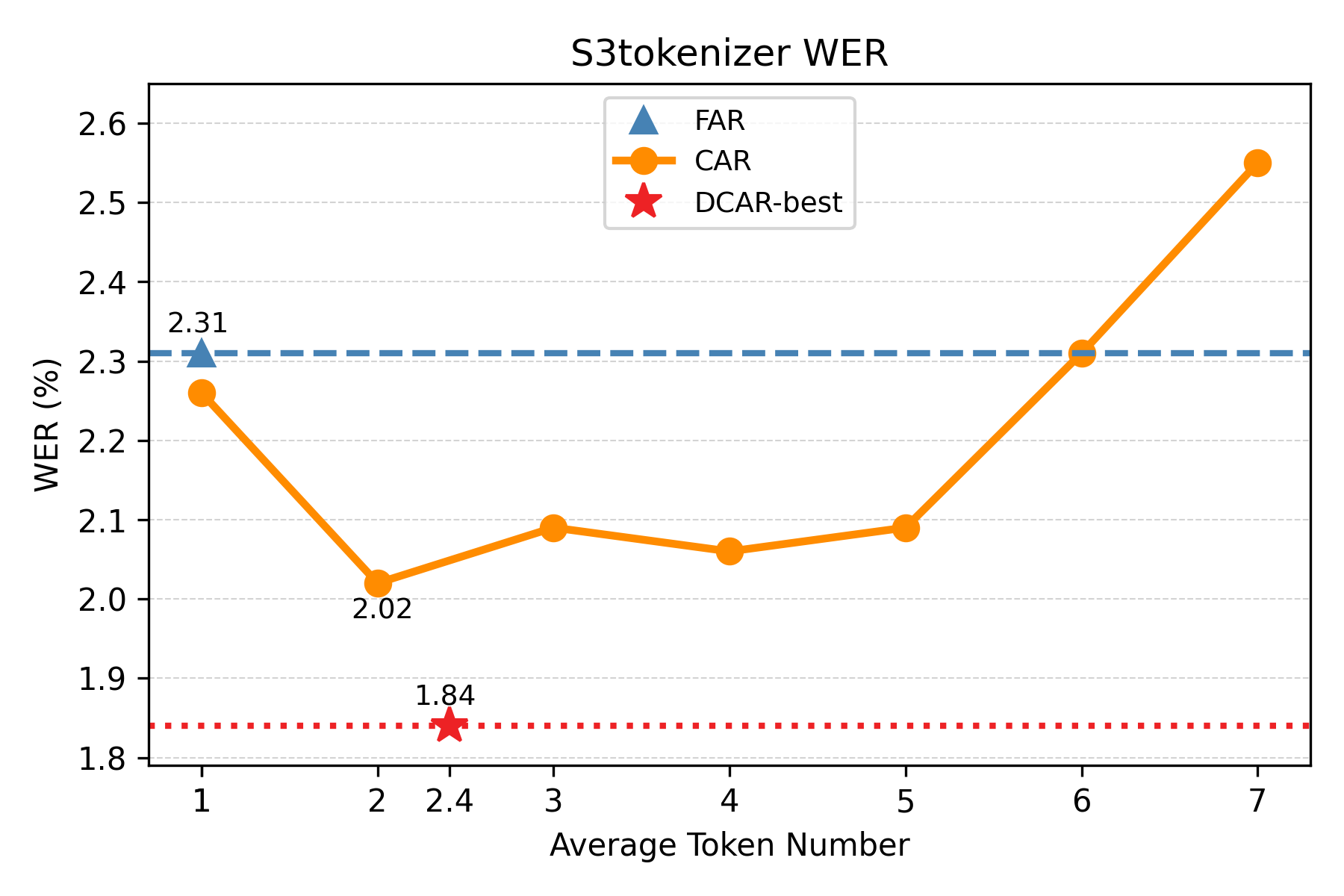}
    \captionof{figure}{}
    \label{fig:s3tokenzier}
\end{minipage}

This section presents an overview of the key evaluation results. As shown in Figures~\ref{fig:hubert} and~\ref{fig:s3tokenzier}, DCAR exhibits both robustness and inference acceleration.

\begin{figure}[ht]
  \centering
  \begin{minipage}[t]{0.5\textwidth}
    \vspace{-100pt}
    \subsection{Scalability}
    \label{E2}
    Experimental results demonstrate the scalability of our method. Specifically, we train the CAR TTS model using HuBERT tokens on the large-scale LibriHeavy dataset (50k hours) for 2 epochs. The results, presented in Figure~\ref{fig:hubert_libriheavy}, highlight the robustness and speed of the synthesis process. For this experiment, the DCPO action guidance range is set to [3,4,5,6].
  \end{minipage}%
  \hfill
  \begin{minipage}[t]{0.45\textwidth}
    \centering
    \includegraphics[width=\linewidth]{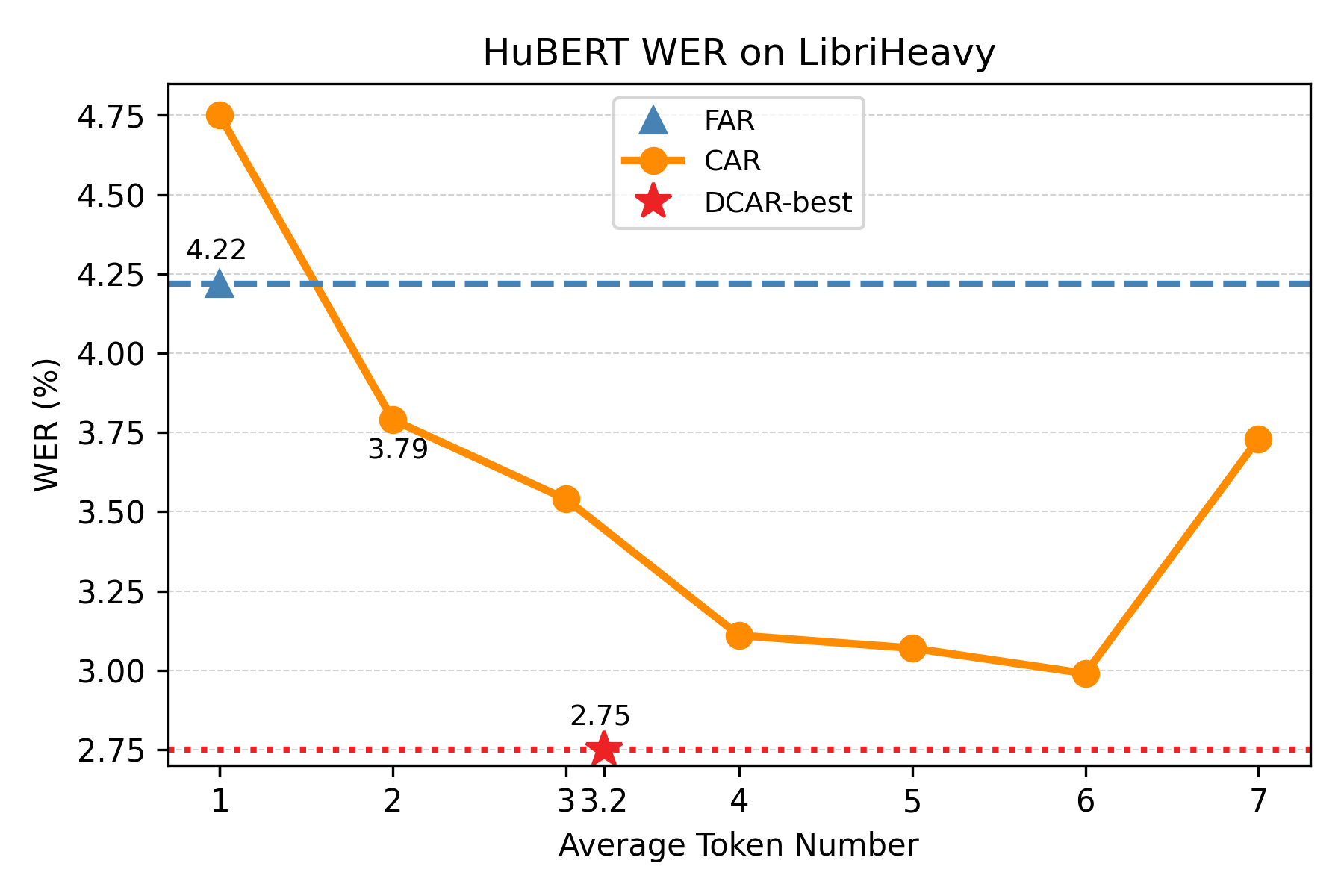}
    \captionof{figure}{Synthesis performance of HuBERT token on LibriHeavy.}
    \label{fig:hubert_libriheavy}
  \end{minipage}
\end{figure}

\subsection{Implementation on CosyVoice}
\label{implementation on CosyVoice}

To validate the generality of our approach, we implement DCAR on the open-source TTS framework CosyVoice. Specifically, we augment the base CosyVoice model with 6 additional heads, train it on LibriTTS for 20 epochs, and train DCPO using the 980-sample tiny set for 5 epochs. The zero-shot TTS results on our test set, shown in Table~\ref{tab:table-cosyvoice}, demonstrate the effectiveness of our method.

\begin{table*}[t]
  \centering
  \small
  \begin{tabular}{l|ccccc}
    \toprule
    \multirow{2}{*}{\textbf{Method}} 
      & \multirow{2}{*}{\textbf{Avg.\ Token}}
      & \multicolumn{3}{c}{\textbf{Quality}} 
      & \textbf{Efficiency} \\
      & 
      & WER$\downarrow$ & SECS$\uparrow$ & UTMOS$\uparrow$ 
      & RTF$\downarrow$ \\
    \midrule

    \multicolumn{6}{c}{\textit{\textbf{CosyVoice-$\mathcal{S}^3$-Tokenizer-50Hz}}} \\
    \midrule
    FAR                     & 1    & 4.74 & 0.819 & 4.25$_{\pm0.017}$ & 0.250  \\
    \midrule
    \multirow{3}{*}{CAR -- WER top3}
                            & 2   & 3.16 & 0.818 & 4.24$_{\pm0.018}$ & 0.184  \\
                            & 3   & 3.73 & 0.819 & 4.21$_{\pm0.020}$ & 0.162  \\
                            & 4   & 3.79 & 0.819 & 4.13$_{\pm0.023}$ & 0.149  \\
    \midrule
    \textbf{\MethodName~-- {[}2,3{]}}   & 2.12   & 2.84 & 0.820 & 4.19$_{\pm0.022}$ & 0.215 \\
    \textbf{\MethodName~-- {[}2,3,4{]}} & 2.01   & 2.93 & 0.818 & 4.19$_{\pm0.021}$ & 0.218 \\
    
    \bottomrule
  \end{tabular}
  \caption{Performance of FAR, CAR, DCAR implemented on CosyVoice. ``Avg. Token'' refers to the average number of speech token simultaneously generated in each step.}
  \label{tab:table-cosyvoice}
  \vspace{-5pt}
\end{table*}

\end{document}